\documentclass[aps,twocolumn,showpacs]{revtex4}
\usepackage{bm}
\usepackage{graphicx}
\usepackage{epsfig}
\begin{document}

\title{Spin photocurrents and circular photon drag effect
in (110)-grown\\ quantum well structures}

\author{V.\,A.\,Shalygin$^{\,\natural}$\footnote{e-mail: shalygin@rphf.spbstu.ru},
H.\,Diehl$^{\,\flat}$, Ch.\,Hoffmann$^{\,\flat}$,
S.\,N.\,Danilov$^{\,\flat}$, T.\,Herrle$^{\,\flat}$,
S.\,A.\,Tarasenko$^{\,\S}$, D.\,Schuh$^{\,\flat}$,
Ch.\,Gerl$^{\,\flat}$, W.\,Wegscheider$^{\,\flat}$,
W.\,Prettl$^{\,\flat}$, and S.\,D.\,Ganichev$^{\,\flat}$}
\affiliation{\vskip 4pt $^\natural$~St.\,Petersburg State
Polytechnic University, 195251 St.\,Petersburg, Russia}
\affiliation{$^\flat$~Fakult\"{a}t Physik, University of
Regensburg, 93040 Regensburg, Germany}
\affiliation{$^\S$~A.\,F.\,Ioffe Physico-Technical Institute,
Russian Academy of Sciences, 194021 St.\,Petersburg, Russia}

\begin{abstract}
We report on the study of spin photocurrents in (110)-grown
quantum well structures. Investigated effects comprise the
circular photogalvanic effect and so far not observed circular
photon drag effect. The experimental data can be described by an
analytical expression derived from a phenomenological theory. A
microscopic model of the circular photon drag effect is developed
demonstrating that the generated current has spin dependent
origin.
\end{abstract}

\pacs{73.50.Pz, 72.25.Fe, 72.25.Rb, 78.67.De}

\maketitle

\section{Introduction}
There is a growing interest in the field of spintronics with the
aim of controlling and manipulating electron spins in
microelectronic devices. A key factor for semiconductor
spintronics is spin relaxation time: it must be sufficiently long
for the processing of information encoded as spin polarization. In
the case of a two-dimensional semiconductor structure  spin
relaxation time strongly depends on the growth direction. It was
shown that in (110)-grown GaAs/AlGaAs quantum wells (QWs) spin
relaxation time is considerably longer compared to that in
(001)-oriented QWs and, therefore,  it can be increased to
nanoseconds  even at room temperature~\cite{Ohno1999}. This is due
to the fact that in (110)-grown QWs the D'yakonov-Perel' mechanism
of spin relaxation dominating in GaAs heterostructures is
suppressed. This observation has attracted a great deal of
attention to spin dependent phenomena in (110)-oriented
structures. An effective access to these phenomena in low
dimensional structures is provided by spin photocurrents like
circular photogalvanic effect
(CPGE)~\cite{Ganichev2001,Bieler05,Yang06} and spin-galvanic
effect (SGE)~\cite{Ganichev2002}, allowing investigation of spin
relaxation times, spin splitting of the band structure, symmetry
properties, etc. (for a review
see~\cite{Ganichev2006,Ch7sturman}). So far spin photocurrents
were mostly studied in (001)- and (113)-grown heterostructures.

In this letter we present investigations of the circular
photogalvanic effect in $n$-type (110)-grown GaAs/AlGaAs QWs and
report on the  observation of a  new effect caused by transfer of
both linear and angular momenta of photons to free carriers. The
latter effect, called circular photon drag effect, was
theoretically predicted in
Refs.~\cite{Ivchenko1980,Belinicher1981}, but so far not observed.
All previous investigations dealt with the linear photon drag
effect where the inversion of the light helicity does not affect
the sign and magnitude of the current (for review
see~\cite{Ganichev2006,Ch7Yaroshetskii80p173,Ch7Gibson80p182}).
The circular photon drag effect reported here, in contrast,
represents the photon drag current which reverses its direction by
inversion of the light helicity from left-handed to right-handed
and vice versa.

\section{Experimental technique}
The experiments are carried out at room temperature on
asymmetrical (110)-oriented GaAs$/$Al$_{x}$Ga$_{1-x}$As molecular
beam epitaxy grown heterostructures containing 100 QWs of 8.2~nm
width separated by 40 nm barriers ($x = 0.34$). Two $n$-type
structures with electron concentration $n_s$ about $7 \cdot
10^{11}$~cm$^{-2}$ per QW and various doping profiles are investigated.
The sample A
contains Si-doped layer of 10~nm width in each
barrier shifted from the barrier center by the distance of 10 nm.
In sample B
the doped layer  of the same width is placed in the
center of each barrier. Samples have sizes about $5\times 5$ mm$^2$.
The sample edges are oriented along
$x\parallel[1{\bar 1}0]$ and $y\parallel[00{\bar 1}]$ in the QW
plane, the $z$-axis points parallel to the structure growth
direction.  To measure
electrical currents ohmic contacts are prepared in the center of
each  sample edge.

The measurements of photocurrents are carried out under excitation
of the samples with infrared or terahertz radiation at normal and
oblique incidence. The source of infrared radiation is  a Q-switch
CO$_2$-laser with operating spectral range (9.2--10.8~$\mu$m)
corresponding to $inter$-subband transitions between the lowest and
the first excited subbands  of the investigated QWs. Pulsed THz radiation is
obtained applying an optically pumped pulsed molecular
laser~\cite{Ganichev2006}.  Several  wavelengths  between 77
and 496~$\mu$m have been selected using NH$_3$, D$_2$O and
CH$_3$F as active media. Terahertz radiation causes the
$intra$-subband (Drude-like) absorption of the radiation. The geometry of
the experiment is sketched in the inset of Fig.~\ref{fig1}(b). The
photocurrent is measured in unbiased structures via the voltage
drop across a $50\: \Omega$ load resistor.

In this work we examine helicity dependent photocurrents,
$J^{circ}_x$, i.e. currents which reverse their sign upon
switching the radiation helicity. In order to extract such a
current contribution from the measured total current we determined
the response to $\sigma_+$ and $\sigma_-$ radiation and evaluated
the data after
\begin{equation}
\label{jcirc} J^{circ}_x =
\left[J_x(\sigma_{+})-J_x(\sigma_{-})\right]/2 \:.
\end{equation}
The right-handed ($\sigma_{+}$) and left-handed ($\sigma_{-}$)
circularly polarized radiation is achieved by means of a Fresnel
rhomb in the infrared and $\lambda$-quarter quartz plates in the
THz range.

\section{Experimental results and discussion}
Irradiating the samples at normal incidence we detected a
photocurrent which is proportional to the radiation helicity
$P_{circ}$ and whose temporal structure reproduces that of the
laser pulse being of the order of 100~ns. This helicity dependent
current has been observed with the contact pairs aligned along
$x\parallel [1\bar{1}0]$ only. All these  features hold for
infrared as well as  THz wavelengths applied and are in agreement
with phenomenological theory. In fact, asymmetric (110)-oriented
heterostructures used in our experiments belong to media of C$_s$
point-group symmetry. In this symmetry the circular photocurrent
density $j_x$ excited by light incident in the $xz$ plane is given
by
\begin{equation}
\label{incidence} j_x = \gamma_{xz}\: t_p t_s \frac{q_z}{q}
{E_0}^2 P_{circ}\:,
\end{equation}
where $\mathbf{\gamma}$ is the second rank pseudo-tensor
describing the sum of the circular photogalvanic
effect~\cite{Ganichev2001} and optical orientation induced
spin-galvanic effect~\cite{Ganichev2002}, $t_p$ and $t_s$ are the
transmission coefficients for $p$ and $s$ components of the light
electric field, $\mathbf{q}$ is  the light wave vector inside the
medium, $E_0$ is the electric field amplitude of the incident
light, and $P_{circ}$ is the light helicity ($P_{circ}=\pm1$ for
$\sigma_{\pm}$ polarization, respectively). The difference of the
currents for $P_{circ}=\pm1$ yields $j^{circ}_x$. The dependence
of the photocurrent on the angle of incidence $\Theta_0$ is given
by $q_z/q=\cos{\Theta}$ and Fresnel's formulas for $t_p$ and
$t_s$, here $\Theta$ is the angle of refraction defined by
$\sin{\Theta} = \sin{\Theta_0}/n_{\omega}$, with $n_{\omega}$
being the index of refraction.

\begin{figure}
\centerline{\epsfxsize 83mm \epsfbox{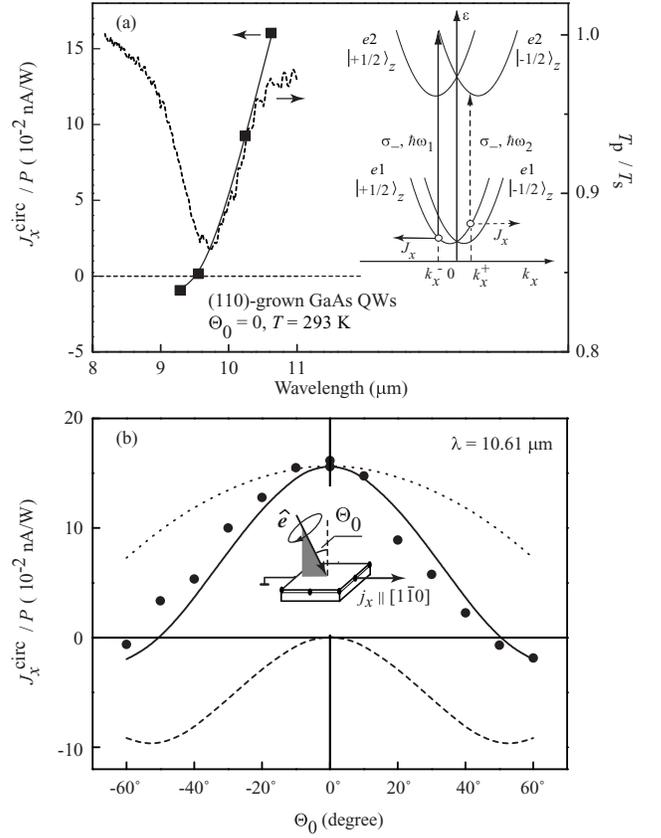}} \caption{(a)
Spectrum of helicity dependent photocurrent $J_x^{circ}$
normalized to the laser power $P$ measured in sample~A illuminated
by infrared laser radiation under normal incidence (squares) and
spectrum of optical transmission ratio $T_p/T_s$ (dashed line) for
light of \textit{p}- and \textit{s}-polarization, respectively.
Solid line is a guide for the eye. The transmission spectrum is
measured under oblique incidence at $\Theta_0=60^{\circ}$. The
inset shows the model of the circular photogalvanic effect caused
by direct intersubband transitions and illustrates the spectral
inversion of the photocurrent induced by $\sigma_-$ radiation due
to reduction of the photon energy from $\hbar \omega_1$ to $\hbar
\omega_2$.
(b) Angular dependence of normalized helicity dependent
photocurrent $J_x^{circ} / P$ obtained for sample~A. The doted,
dashed, and solid curves are fit after Eq.~(\protect
\ref{jcircdrag}) and represent, correspondingly, the term in
square brackets, the last term on the right hand side of
Eq.~(\protect \ref{jcircdrag}), and the sum of both terms. The
inset shows the experimental geometry.} \label{fig1}
\end{figure}

Measurements of spectral behavior of $J_x^{circ}$ in the infrared
range show a spectral inversion as plotted in Fig.~\ref{fig1}(a).
The point of the inversion corresponds to the maximum of resonant
intersubband absorption which is measured by means of the Fourier
transform transmission spectroscopy. The fact that the
photocurrent changes sign by tuning the wavelength indicates that
it is mainly caused by the CPGE outweighing the spin-galvanic
effect~\cite{Ganichev2003sge}. The model picture of the CPGE
illustrating the spectral sign inversion of the current at the
center of the absorption line is sketched in the inset of
Fig.~\ref{fig1}(a) after~\cite{Ganichev2003}.
%
In structures of C$_s$ symmetry the spin-orbit coupling splits the
electron spectrum into spin branches with the spin components
$s_z=\pm 1/2$ along the growth direction. The relevant
contribution to spin-orbit part of the effective Hamiltonian is
given by $\beta_{zx} \sigma_{z} k_x$, where $\beta_{zx}$ is a
parameter and $\sigma_z$ is the Pauli matrix. Due to the optical
selection rules, the normal-incidence circularly polarized
radiation, e.g. $\sigma_-$, induces direct optical transitions
from the subband $e1$ with the spin $s_z=+1/2$ to the subband $e2$
with $s_z=-1/2$. Monochromatic radiation with the certain photon
energy, say $\hbar \omega_1$, induces the transitions only at a
fixed wave vector $k_x^-$ where the photon energy matches the
transition energy as indicated by the solid vertical arrow in the
inset of Fig.~\ref{fig1}(a). Thus, the intersubband excitation
results in an imbalance of the momentum distribution between
positive and negative $k_x$ in both subbands yielding an electric
current. As in our QWs the energy separation between the subbands
$\varepsilon_{21}$ is larger than the energy of longitudinal
optical phonons ($\varepsilon_{21} \approx 100$~meV,
$\hbar\Omega_{\rm LO} = 35$~meV), the nonequilibrium distribution
of electrons in the subband $e2$ relaxes rapidly due to the
emission of phonons. By that the contribution of the subband $e2$
to the electric current vanishes. Therefore, the magnitude and the
direction of the current, shown in the inset of Fig.~\ref{fig1}(a)
by the solid horizontal arrow, is determined by the group velocity
and the momentum relaxation time $\tau_{e1}$ of the photogenerated
``holes'' in the subband $e1$ with $s_z=+1/2$. Obviously, the
whole picture mirrors and the current direction reverses by
switching the circular polarization from left- to right-handed.
Spectral inversion of the photocurrent at fixed helicity also
follows from this model picture. Indeed, as is shown in
Fig.~\ref{fig1}(a), decreasing the photon frequency to
$\hbar\omega_2$ shifts the transitions toward positive $k_x$
(dashed vertical arrow), and the direction of the current reverses
(dashed horizontal arrow). The inversion of the current direction
takes place at the photon energy $\hbar \omega_{\rm inv}$
cor\-res\-pon\-ding to the optical transitions from the spin
subband minima. This mechanism is based on spin splitting due to
$\sigma_z k_x$ terms and predicts, in accordance with the
phenomenological equation~(\ref{incidence}), that the current
reaches a maximum at normal incidence and becomes smaller under
oblique incidence keeping the same direction.

\begin{figure}
\centerline{\epsfxsize 83mm \epsfbox{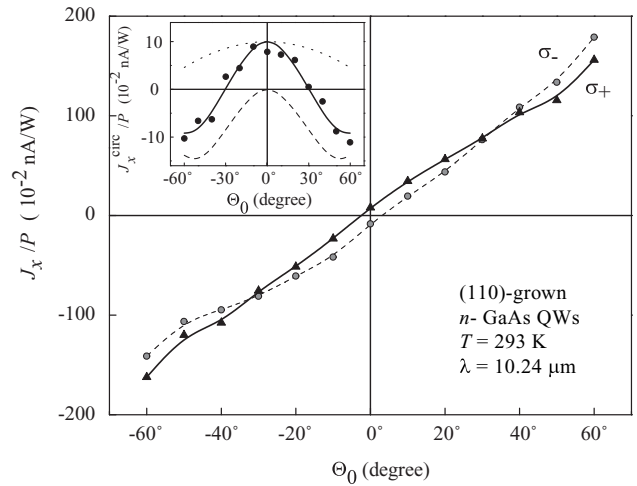}}
\caption{Dependencies of the total photocurrent $J_x$ normalized
to the laser power $P$ on the angle of incidence $\Theta_0$
obtained for sample~B illuminated with right-handed ($\sigma_{+}$,
triangles) and left-handed ($\sigma_{-}$, circles) circularly
polarized radiation. Lines are a guide for the eye. The inset
shows the angular dependence of the helicity dependent part of the
normalized photocurrent, $J_x^{circ} / P$, for the same structure.
The doted, dashed, and solid curves are fit after Eq.~(\protect
\ref{jcircdrag}) and represent, correspondingly, the term in
square brackets, the last term on the right hand side of
Eq.~(\protect \ref{jcircdrag}), and the sum of both terms.}
\label{fig2}
\end{figure}

In order to cross-check the validity of the phenomenological
Eq.~(\ref{incidence}) with respect to our measurements, we
investigated the current response as a function of the incidence
angle $\Theta_0$. In the whole THz range, where the photocurrent
is caused by Drude absorption, we found that the data are well
described by this equation. However, in the infrared range a
qualitative discrepancy to Eq.~(\ref{incidence}) is observed. In
contrast to the sign conserving behavior of the photocurrent given
by $t_p t_s \cos \Theta$, the signal in the sample~A changes its
sign twice at $\Theta_0 \approx \pm 50^{\circ}$, see
Fig.~\ref{fig1}(b). The experiment carried out on the sample~B
gave the effect even more pronounced: here the inversion takes
place  at $\Theta_0 \approx \pm 30^{\circ}$, see inset of
Fig.~\ref{fig2}. We note that in contrast to normal incidence of
radiation, where only helicity dependent current is observed, at
oblique incidence a substantial helicity independent contribution
to the total current is found, see Fig.~\ref{fig2}. This
contribution is due to the linear photogalvanic effect
~\cite{Magarill82,Gusev87} and the  linear photon drag
effect~\cite{Luryi87,Sigg90}, which are out of scope of this
letter. Nevertheless, the helicity dependent contribution
$J_x^{circ}$ is large enough and is easily measurable. In
particular, the twofold sign inversion of the helicity dependent
photocurrent with incidence angle variation is clearly seen in
Fig.~\ref{fig2}.

This angle inversion of the current direction cannot be explained
in the framework of the conventional theory of the CPGE or
optically excited SGE which ignores the linear momentum transfer
from photons to free carriers. Taking into account the linear
momentum of the photon, neglected in Eq.~(\ref{incidence}), we
obtain an additional contribution to the current excited by
circularly polarized light. Then, the total helicity dependent
photocurrent in structures of $C_s$ symmetry is given by
\begin{equation}
\label{dragcirc}
j_x ={t_pt_s}\left\{ \left[ \left( \gamma_{xz} +
q_z T_{xzz} \right) \frac{q_z}{q} \right] + q_x T_{xxx}
\frac{q_x}{q} \right\} {E_0}^2 P_{circ}\:,
\end{equation}
where $\mathbf{T}$ is the third rank tensor which describes the
circular photon drag effect. Following Eq.~(\ref{dragcirc}) one
obtains the angular dependence of the photocurrent
\begin{eqnarray}
\label{jcircdrag}
j_x &=& t_p t_s \left\{\left[ \left( \gamma_{xz}
+ q T_{xzz} \cos
\Theta \right) \cos \Theta \right]  \right. \nonumber\\
&&+ \left. q T_{xxx} \sin^2 \Theta \right\}{E_0}^2 P_{circ} \:.
\end{eqnarray}

Equation~(\ref{jcircdrag}) shows that  the circular photon drag
effect given by terms containing the linear photon momentum {\bf
q} can be observed, in principle, at both normal ($\Theta=0$) and
oblique incidence. However, distinction between contributions of
the CPGE and of the circular photon drag effect for $\Theta = 0$
is not an easy task. It may be done keeping in mind that the
replacement $P_{circ}\rightarrow -P_{circ}$ and $q_z \rightarrow -
q_z$ in Eq.~(\ref{dragcirc}) conserves the first term in the
square brackets on the right hand side of Eq.~(\ref{dragcirc})
while changes the sign of the second term. Experimentally it can
be realized putting a mirror behind the sample and comparing the
current magnitudes with and without mirror. However, such a method
requires very high accuracy of adjustment.

Much more reliable access to the circular photon drag effect is
provided by studying the angular dependence of the photocurrent.
Indeed, the terms in square brackets in Eq.~(\ref{jcircdrag}) have
a maximum at normal incidence and their contribution to the
current decreases with increasing the angle of incidence. At the
same time the circular photon drag effect given by the last term
in Eq.~(\ref{jcircdrag}) vanishes at normal incidence and
increases with $\left| \Theta_0 \right|$. This interplay of the
current contributions may result in the observed twofold sign
inversion of the total current by the variation of $\Theta_0$ from
$-\pi/2$ to $\pi/2$ if the circular photon drag and the CPGE
photocurrents are oppositely directed. The fits of
Eq.~(\ref{jcircdrag}) to the experimental data for both QW
structures are shown in Fig.~\ref{fig1}(b) and in the inset of
Fig.~\ref{fig2}. The plotted curves represent the terms in square
brackets (doted curves), the last term on the right hand side of
Eq.~(\ref{jcircdrag}) (dashed curves), and the sum of both terms
(solid curves). To fit the data for each sample we use an ordinate
scaling parameter for the doted curve to obtain agreement at
normal incidence, where the last term on the right hand side of
Eq.~(\ref{jcircdrag}) vanishes. Then, the dashed curve is scaled
to fit the data in the whole range of the incidence angles
$\Theta_0$. It is seen that the phenomenological
equation~(\ref{jcircdrag}) describes well the experimental angular
dependence of the photocurrent. The contribution to the circular
photon drag effect given by the component $T_{xxx}$ reaches its
maximum in GaAs/AlGaAs structures at $\Theta_0 \approx \pm
50^\circ$.

\section{Microscopical model}
Now we discuss the microscopic picture of the observed  circular
photon drag effect given by the last term on the right hand side
of Eq.~(\ref{jcircdrag}). The tensor $\mathbf{T}$ is not invariant
under time inversion. Therefore, dissipative processes should be
involved in the microscopic model of the effect. The proposed
model includes three stages.

\begin{figure}
\centerline{\epsfxsize 85mm \epsfbox{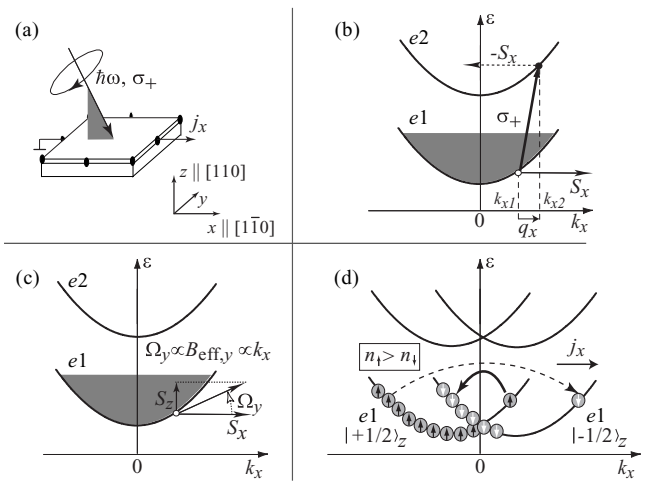}} \caption{(a)
Geometry of the experiment. (b)-(d) Sketch of three sequential
stages of the microscopical model of the circular photon drag
effect: (b) helicity and photon wave vector dependent
photoexcitation, (c) spin rotation in an effective magnetic field
caused by spin-orbit coupling, and (d) asymmetrical spin
relaxation resulting in an electric current flow due to the
spin-galvanic effect.} \label{fig3}
\end{figure}

The first stage is a helicity and photon wave vector dependent
photoexcitation. The intersubband absorption of circularly
polarized radiation is a spin dependent process. While at normal
incidence the absorption of circularly polarized light is due to
spin-flip processes (see inset in Fig.~\ref{fig1}(a)), under
oblique excitation due to selection rules the absorption is
dominated  by spin conserving transitions~\cite{Ivchenko2004}.
However, the rates of these spin conserving transitions are
different for electrons with the spin oriented parallel and
antiparallel to the in-plane direction of light propagation ($x$
in Fig.~\ref{fig3}(a)). In Fig.~\ref{fig3}(b) the dominating
optical transitions are sketched by an inclined arrow to take into
account the linear momentum of the photon involved. As a result of
the linear momentum transfer the optical transitions occur at a
distinct initial electron wave vector determined by energy and
momentum conservation. The angular momenta of photons yield a spin
polarization $S_x$ at $k_{x1}$ and $-S_x$ at  $k_{x2}$ in the
subbands $e1$ and $e2$, respectively. These spin polarizations are
indicated in Fig.~\ref{fig3}(b) by solid and dashed horizontal
arrows. While optical excitation results in a spin polarization at
well determined wave vectors, $k_{x1}$ in the subband $e1$ and
$k_{x2}$ in $e2$, the electrons in the upper subband have
sufficient energy to emit optical phonons and rapidly relax due to
this process. Thus, the spin polarization $S_x$ in the lower
subband only is connected with electrons with the well defined
momentum ($k_{x1}$ in Fig.~\ref{fig3}(b)).

The second stage is spin precession in an effective magnetic field
caused by the Rashba or the Dresselhaus spin-orbit coupling. The
orientation and the strength of this effective magnetic field is
determined by the direction and the magnitude of the electron wave
vector. As our optical excitation results in the spin polarization
$S_x$ of electrons with the certain wave vector $k_{x1}$, the
effective magnetic field linked to this wave vector acts on the
electron spin. Spins of the electrons, directed just after
photoexcitation along the $x$ axis, precess in the effective
magnetic field which has both $\Omega_z \propto k_x$ and $\Omega_y
\propto k_x$ components. As a consequence of the precession the
spin components $S_y$ and $S_z$ appear, see Fig.~\ref{fig3}(c) for
component $S_z$. Under steady-state excitation the generation
rates of the spin components $S_y$ and $S_z$ are determined by the
average angle of spin rotation in the effective magnetic field.

In the third stage, the nonequilibrium spin polarization $S_z$
obtained in the first two stages of the proposed model description
drives an electric current. This is due to the spin-galvanic
effect caused by asymmetric spin relaxation \cite{Ganichev2002}.
The mechanism is briefly sketched in Fig.~\ref{fig3}(d) where we,
like in the inset of Fig.~\ref{fig1}(a), take into account the
spin-orbit splitting of the subbands due to $\sigma_z k_x$-terms
in the effective Hamiltonian. The difference in carrier
populations in the spin branches $s_z=\pm1/2$ of the ground
subband 
($n_\uparrow > n_\downarrow$) causes spin relaxation. The rate of
spin-flip scattering depends on the electron wave vectors in the
initial and final states that is illustrated by bent arrows of
different thicknesses. The transitions of different rates lead to
an asymmetric distribution of electrons within each spin branch.
As a result an electric current $j_x$ arises. The symmetry
analysis shows that the relaxation of the spin component $S_y$ is
also accompanied by generation of an electric current along the
$x$ direction.

The process of the third stage, spin-galvanic effect, was already
studied in~\cite{Ganichev2002,Ganichev2003sge,Ivchenko1990}.
Therefore, we concentrate below on the first two stages and
consider them as a specific kind of optical orientation of
electron spins, which is caused by simultaneous transfer of photon
linear and angular momenta to the carriers.

The intersubband light absorption in $n$-doped QW structures is a
resonant process and possible if the photon energy equals the
energy spacing between the subbands. In the single-band
approximation, direct optical transitions from the subband $e1$ to
the subband $e2$ conserve spin orientation and can be induced only
under oblique incidence of the light with nonzero $p$ component of
polarization. These selection rules are violated if one takes into
account $\mathbf{k}$$\cdot$$\mathbf{p}$ admixture of the
valence-band states to the conduction-band wave functions. In this
model the light of both $s$- and $p$-polarization can induce
intersubband optical transitions, and the transitions become spin
dependent~\cite{Ganichev2003,Ivchenko2004}. We assume that
electrons occupy the ground subband $e1$ and the size-quantization
energy is substantially larger than the mean kinetic energy in the
QW plane. Then, the spin matrix of electron photogeneration in the
subband $e1$ has the form
\begin{equation}\label{G_k}
G_{\mathbf{k}} = - \frac{2\pi}{\hbar}  M^{\dag} M f_{\mathbf{k}}
\, \delta(\hbar\omega + \varepsilon_{1,\mathbf{k}} -
\varepsilon_{2,\mathbf{k}+\mathbf{q}_{\parallel}}) \:,
\end{equation}
where $M$ is a $2\times2$ matrix describing the intersubband
optical transitions, $M^{\dag}$ is the hermitian conjugate matrix,
$f_{\mathbf{k}}$ is the function of equilibrium carrier
distribution, $\varepsilon_{1,\mathbf{k}}=\hbar^2 k^2/2m^*$ and
$\varepsilon_{2,\mathbf{k}}=\varepsilon_{21}+\hbar^2 k^2/2m^*$ are
the electron dispersions in the subbands $e1$ and $e2$,
respectively, $m^*$ is the effective electron mass,
$\varepsilon_{21}$ is the energy spacing between the subbands, and
$\mathbf{q}_{\parallel}$ is the in-plane component of the photon
wave vector. The $\delta$-function in Eq.~(\ref{G_k}) reflects the
resonant behavior of the intersubband optical transitions. In real
QW structures the spectral width of the resonance is broadened due
to finite scattering time of carriers, fluctuations of the QW
width, etc.  To describe the broadening one can replace the
$\delta$-function by a normalized function $\delta_{\Gamma}$ which
corresponds to the absorption spectrum in the real structure. To
first order in the $\mathbf{k}$$\cdot$$\mathbf{p}$ theory, the
matrix $M$ is given by~\cite{Ivchenko2004}
\begin{equation}\label{eq18}
M = -\frac{eA}{cm^*}\,p_{21}\left[
\begin{array}{cc} e_z & \Lambda (e_x - i e_y) \\ - \Lambda (e_x + i
e_y) & e_z
\end{array} \right]\;,
\end{equation}
where $A$ is the amplitude of the electro-magnetic wave related to
the light intensity by $I=A^2\omega^2 n_{\omega} / (2 \pi c)$, $c$
is the light velocity, and $p_{21}$ is the momentum matrix element
between the envelope functions of size quantization
$\varphi_{1}(z)$ and $\varphi_{2}(z)$ in the subbands $e1$ and
$e2$,
\begin{equation}
p_{21} = - i \hbar \int \varphi_{2}(z) \frac{\partial}{\partial z}
\varphi_{1}(z)\: dz \:.
\end{equation}
The parameter $\Lambda$ originates from ${\bf k}$$\cdot$${\bf p}$
admixture of valence-band states to the electron wave function and
is given by
\begin{equation} \Lambda = \frac{ \varepsilon_{21} \Delta (2
E_g + \Delta)}{2E_g (E_g + \Delta)(3 E_g + 2 \Delta)}\:,
\end{equation}
where $E_g$ is the energy of the band gap, and $\Delta$ is the
energy of spin-orbit splitting of the valence band.

Absorption of circularly polarized light leads to spin orientation
of photoexcited carriers. We assume that the momentum relaxation
time $\tau_{e1}$ is shorter than the precession period in the
effective magnetic field, $\Omega \tau_{e1}\ll 1$. Then, the spin
generation rate in the subband $e1$ has the
form~\cite{Tarasenko2005}
\begin{equation}\label{S_dot}
\dot{\mathbf{S}} = \sum_{\mathbf{k}} \mathbf{g}_{\mathbf{k}} +
\sum_{\mathbf{k}} \tau_{e1}
[\mathbf{\Omega}\times\mathbf{g}_{\mathbf{k}}] \:,
\end{equation}
where
$\mathbf{g}_{\mathbf{k}}=\mathrm{Tr}(\bm{\sigma}G_{\mathbf{k}})/2$
is the rate of spin photogeneration into states with the wave
vector $\mathbf{k}$, $\bm{\sigma}$ is the vector of the Pauli
matrices. The first term in Eq.~(\ref{S_dot}) describes optical
orientation of carriers in the moment of photoexcitation, while
the second term stands for spin orientation, which is caused by
spin dependent asymmetry of excitation in $\mathbf{k}$-space
followed by spin precession in the effective magnetic field. It is
the term that describes optical orientation by circularly
polarized light, which is related to the transfer of photon linear
momenta to charge carriers and vanishes if
$\mathbf{q}_{\parallel}=0$.

In asymmetrically (110)-grown QW structures the Larmor frequency
corresponding to the effective magnetic field has the form
\begin{equation}
\mathbf{\Omega} = \frac{2}{\hbar} (\beta_{xy}k_y, \,
\beta_{yx}k_x, \, \beta_{zx}k_x) \:,
\end{equation}
where $\beta_{xy}$, $\beta_{yx}$ and $\beta_{zx}$ are constants of
the spin-orbit interaction. As in the experiment described above,
we consider that the light wave vector $\mathbf{q}$ lies in the
$xz$ plane. Then, for the Boltzmann distribution of carriers, one
derives
\begin{equation}\label{S_x}
\dot{S}_x =  \Lambda \eta_z(\hbar\omega) \frac{q_x}{2q} \frac{I
P_{circ}}{\hbar\omega} \:,
\end{equation}
\begin{equation}\label{S_y}
\dot{S}_y = -   q_x \, \bar{\varepsilon} \, \Lambda
\frac{\beta_{zx}\tau_{e1}}{\hbar} \frac{q_x}{q} \frac{d\,
\eta_z(\hbar\omega)}{d \,\hbar\omega} \frac{I
P_{circ}}{\hbar\omega} \:,
\end{equation}
\begin{equation}\label{S_z}
\dot{S}_z = \left[ \eta_z(\hbar\omega) \Lambda^2 \frac{q_z}{2q} +
q_x \bar{\varepsilon} \, \Lambda \frac{\beta_{yx}\tau_{e1}}{\hbar}
\frac{q_x}{q} \frac{d\, \eta_z(\hbar\omega)}{d \,\hbar\omega}
\right] \frac{I P_{circ}}{\hbar\omega} \:,
\end{equation}
where $\bar{\varepsilon}=k_B T$ is the mean kinetic energy of
equi\-libri\-um carriers, $T$ is the temperature,
$\eta_z(\hbar\omega)$ is the QW absorbance for light polarized
along the growth direction,
\begin{equation}
\eta_z (\hbar\omega) =\frac{4\pi^2 \alpha}{n_{\omega}}
\,\frac{\hbar |p_{21}|^2}{m^{*2}\omega} \,n_s
\,\delta_{\Gamma}(\hbar\omega-\varepsilon_{21}) \:,
\end{equation}
and $\alpha$ is the fine-structure constant.

As addressed above, relaxation of spin components $S_y$ and $S_z$
in (110)-oriented QW structures is accompanied by the generation
of an electric current along the $x$ axis due to the spin-galvanic
effect. Equations~(\ref{S_y}) and (\ref{S_z}) show that spin
components $S_y$ and $S_z$ contain contributions proportional to
the photon wave vector $q_x$ and the light helicity $P_{circ}$.
Therefore, the generated photocurrent can be gathered in a class
of photon drag effects denoted as the circular photon drag effect.

\section{Conclusion}
We have studied spin dependent photocurrents  in $n$-doped
zinc-blende-based (110)-grown QWs caused by  direct intersubband
transitions induced by infrared radiation and the Drude absorption
of terahertz radiation. The helicity dependent photocurrent
observed at normal incidence of the infrared radiation is mostly
due to the circular photogalvanic effect. Under oblique incidence,
however, the helicity dependent photocurrent at large angles of
incidence flows in the direction opposite to that excited at
normal incidence. We demonstrated that the inversion of the
current sign is a result of the interplay between the circular
photogalvanic effect and the circular photon drag effect.
Microscopic theory of the latter effect developed in the present
work is based on optical spin orientation sensitive to the photon
wave vector and subsequent asymmetric spin relaxation.

We are grateful to E.L.~Ivchenko, L.E.~Golub, and V.V.~Bel'kov for
helpful discussions and thank I. Gronwald for technical
assistance. This work was supported by the Deutsche
For\-schungs\-ge\-mein\-schaft through SFB 689 and GRK638, the
RFBR, programs of the RAS and Russian Ministry of Education and
Science, Russian Science Support Foundation, and Foundation
``Dynasty''-ICFPM.

\end{document}